\documentclass[12pt]{article}

\usepackage{graphicx}
\usepackage{subfigure}
\usepackage{amsmath}
\usepackage{amsfonts}
\usepackage{amssymb}
\usepackage[T1]{fontenc}
\usepackage{calligra}

\def\mpl{{M_{\rm{Pl}}}}

\def\beq{\begin{eqnarray}}
\def\eeq{\end{eqnarray}}
\def\lsim{\mathrel{\rlap{\lower3pt\hbox{\hskip0pt$\sim$}}
     \raise1pt\hbox{$<$}}}         
\def\gsim{\mathrel{\rlap{\lower3pt\hbox{\hskip0pt$\sim$}}
     \raise1pt\hbox{$>$}}}         

\newcommand{\comment}[1]{}

\begin{document}

\begin{flushright}

{NYU-TH-10/07/14}
\end{flushright}

\vskip 1cm

\begin{center}

{\Large {The Big Constant Out, The Small Constant In}}

\end{center}

\thispagestyle{empty}

\vskip 0.5 cm

\begin{center}

{\large  { Gregory Gabadadze}}

 \vspace{.2cm}

{\it Center for Cosmology and Particle Physics,
Department of Physics,}
\centerline{\it New York University, New York,
NY, 10003, USA}

\end{center}

\vskip 1cm

\begin{abstract}

 Some time ago,  Tseytlin has made an original  and unusual proposal  for an action 
 that eliminates  an arbitrary   cosmological constant.  The form of the proposed  action, 
however, is strongly modified by gravity loop effects, ruining  its benefit. 
Here I discuss an embedding of Tseytlin's action into a broader context,  
that enables  to  control the  loop effects. The  broader context 
is another universe, with its own metric and dynamics,  but only globally 
connected to ours.  One possible  Lagrangian for the  other universe 
is that of unbroken  AdS supergravity.   A vacuum energy in our universe 
does not produce any curvature for us,  but  instead increases or decreases the AdS 
curvature  in the  other  universe.  I comment on how to introduce the accelerated  expansion in 
this framework in a technically natural way,  and consider the case where this is done  
by  the self-accelerated solutions of massive gravity and its extensions.

\end{abstract}

\newpage


{\it 1. Introduction:} Nearly a quarter-century ago,   Tseytlin \cite {Tseytlin}  has  proposed 
an approach to the old cosmological constant problem,  using an original 
idea by Linde \cite {Linde}, and certain string-theory developments of that time. 
The proposal is technically well-framed, while 
a highly unconventional  nature of this approach is commensurate with the 
magnitude and longevity of the problem, hence suggesting the approach 
may have a chance of being viable.  

While the proposed  action of \cite {Tseytlin}  enables one to eliminate an arbitrary cosmological constant,
the action itself was argued to be  unstable with respect to quantum corrections,  therefore making the proposal  
not  workable in its original form (see the note added in  \cite {Tseytlin}). The goal of this work is to   
extend  the  proposal  to avoid the quantum loop problem, and to incorporate the dark energy component 
into the  theory in a technically natural way.

\vskip 0.2cm

{\it 2. Tseytlin's proposal:} To set the conventions, consider the action:
\beq
S= \int d^4 x  \sqrt {g} \left ( {1  \over 16 \pi G_N} R  +  L  (g, \psi_n) \right ) \,,
\label{action} 
\eeq
where
$\psi_n\,~n=0,1,2,3...$,  denote all  fields of the theory beyond 
the metric field $g_{\mu\nu}$. Ten Einstein equations 
can be decomposed as  nine traceless  and one trace equation:
\beq
R_{\mu\nu} - \, {1\over 4} g_{\mu\nu} R  = T_{\mu\nu }  - {1\over 4} g_{\mu\nu} T \,, ~~~~~~~~
R+\, T =0 \,.
\label {Einstein}
\eeq
(Unless $G_N$ or $\mpl$  are displayed explicitly, we use  the $ 8 \pi G_N=1$ units).

Instead of these equations,  Tseytlin introduced a   system  where the trace  equation is modified 
(see  the corresponding action  below in eq.  (\ref {Sbar})):
\beq
R_{\mu\nu} - \, {1\over 4} g_{\mu\nu} R  = T_{\mu\nu }  - {1\over 4} g_{\mu\nu} T \,, ~~~~~~~~
R+\, T   = \langle T \rangle -  2 \langle g^{\mu\nu}   {\partial L  \over \partial g^{\mu\nu}}  \rangle\,,
\label{Tseytlin}
\eeq
where  $\langle  \cdots  \rangle $  denotes  a certain space-time average defined as follows:
\beq
\langle  \cdots \rangle  \equiv  { \int d^4 x\sqrt{g} (\cdots )  \over 
  \int d^4 x\sqrt{g}  } \equiv  
  {  [  \cdots ] \over  V_g   }\,.
\label{lrangle}
\eeq 
The modification of the trace equation in (\ref {Tseytlin})  is dramatic: 
local observables on the l.h.s.\footnote{Throughout the paper we use  commonly 
accepted acronyms: "l.h.s."  and "r.h.s.",  for  left and right  hand side 
respectively,  "w.r.t." for with respect to,  "UV" for ultraviolet, and "1PI" for 
1-particle irreducible.} are affected by space-time averaged 
quantities, where the averaging is done over the past and future. 
These averages, when nonzero, have pre-notion of future.
In that sense, this is an acausal modification.  Somewhat similar, but essentially 
different  proposal was made in \cite {ADDG};  a subtle issues 
of defining the average, where  there are more than one vacua,  was also 
raised there.  To begin with, we  envision   a simple universe   
evolving in one vacuum state, and comment on  possible generalizations  later. 
  
If  $V_g \to \infty$, as in Tseytlin's approach, 
then for most of the stuff in the universe the r.h.s. of the trace equation in (\ref {Tseytlin})  
is zero: For any observable, ${\cal O}$,  that is localized 
either in space or in time, the  average $\langle {\cal O} \rangle $   is  zero 
due to the volume factor suppression.  Hence, the  acausality  of the 
trace equation does not manifest itself in the dynamics   of most of the stuff in the universe. 
On the other hand, for a constant Lagrangian, $L=c$ , the r.h.s.  of the trace equation (\ref {Tseytlin})  
is proportional to the constant $c$ itself, and the latter  subtracts the equivalent 
part on the l.h.s., hence leaving the equation independent of $c$!   Therefore,   the main
consequence  of the acausality might be that we don't observe the big cosmological constant in our 
universe \cite {ADDG}.
 
For a scalar field the Lagrangian can be decomposed into  the part  that depends on the metric 
(derivative terms, e.g., the kinetic term)  and the one that is independent  of $g_{\mu\nu}$: 
\beq
L =  L_g (g, \psi_n)  - V(\psi_n)\,.
\label{L}
\eeq
Simplest examples of $V$ are the vacuum energy term $E_{vac}^4$,  scalar mass term $m_\phi^2 \phi^2$, 
scalar potential  $\lambda\phi^4$, or a  linear combination of the above.
The  vacuum energy,  or a constant part of a potential  $V$, 
would give rise to a nonzero  average,  $\langle V \rangle  = [ Const.]/V_g = Const. $ 
Thus,  this quantity would be subtracted  from  the trace $T$ in 
(\ref {Tseytlin}). This is equivalent to  the elimination of the  cosmological constant! 

The fact that a constant term in $L$  is irrelevant,  can also be seen
by looking at the action 
\beq
{\bar S} = {S \over V_g}\,,
\label{Sbar}
\eeq 
that Tseytlin introduced \cite {Tseytlin} as an object which 
has to be varied w.r.t.  $g^{\mu\nu}$ to get  the equations  of (\ref {Tseytlin}).
Any constant shift,  $L \to L+ c$,   gives rise to a  shift 
of the new action by the same constant, ${\bar S} \to {\bar S} +c $,  
that does not affect the equations of motion.

Furthermore, if the potential has  two  minima, one "false" and one "true",
then what is being  subtracted is the value in  the "true" minimum, assuming 
that a transition from  "false" to "true" is possible   in finite time 
in the standard General Relativity  context.  Generically, what is being eliminated is what would have been 
the asymptotic future  value of the vacuum energy density  in GR, as discussed in detail in  \cite {ADDG}.
  
As to the second term on the r.h.s. of the  trace equation in (\ref {Tseytlin}),
it contains only the $L_g$ part of the Lagrangian (\ref {L}); for homogeneous scalar fields
this part eventually decays  on  solutions for which the field settles in its minimum, 
therefore its average $\langle \cdots \rangle$ is zero. Thus, inflation
would generically proceed in a conventional way,  except the phenomenon  
of self-replications is not  straightforward  to incorporate  in this framework \cite {Linde,Linde1}. 

\vskip 0.2cm

{\it 3. Problems with the loops:} While the above approach appears to solve the big 
cosmological constant problem,  at least in the limited context specified above,  there are two  
important issues  that it fails to address: 

First, as  mentioned already in \cite {Tseytlin}, 
the loop corrections should  be  problematic, and they are indeed. 
They  strongly renormalize the form of the action (\ref {Sbar}), and thus 
ruin the solution of the cosmological constant problem.  
Can the issue of the loops be  resolved, by perhaps extending  the proposal?

Second, Tseytlin's mechanism eliminates entirely the cosmological constant. 
Later on,  it  was discovered  that the expansion of the universe  is accelerating \cite{accel}. 
This acceleration can be accounted for by  some form of dark  energy, with  an 
equation of state  parametrized by  $w= -1$.  A cosmological constant has 
precisely that equation of state. Then, the question arises: if one eliminates 
the cosmological constant how does one get to retain dark energy with  $w= -1$?
We'll discuss  how  the accelerated expansion  can be accommodated in this scheme 
in a technically natural way;  one option is to invoke massive gravity  for this purpose.

We proceed in this work by discussing   the  path integral formulation  of the theory
more explicitly.  This  requires an introduction of a special algorithm for  
path integral quantization of  (\ref {Sbar}).  As a result, we'll end  up with two  
different path integrals: one for all non-gravitational interactions quantized with the 
Planck's constant $\hbar$, and another path integral  for gravity quantized with a 
different, dynamically determined Planck's constant.

In the absence of gravity --  in the $\mpl \to \infty$ limit -- $\bar S$ 
differs from the standard action by an overall  $1\over \infty$  factor; the latter   
is field independent,   and thus can be rescaled away \footnote{All  four-volume infinities 
throughout the paper are assumed to be first regularized to yield  finite quantities, 
and the regulator removed only after the equations of motion are derived.}.
Thus, in this limit one would quantize the theory (\ref {Sbar}) 
in a conventional way.   When  dynamical  gravity is included, however,   one needs to  specify  
rules of quantization.  One would not immediately worry about  a UV completion, 
but a low-energy  effective field theory quantization of gravity 
should certainly be a matter of concern:  The Einstein gravity is  a good low-energy 
effective quantum field theory  below the Planck energy scale \cite {Donoghue},    and 
any of its extension  should strive to retain this virtue,  below a certain energy scale. 
It will  be our  goal  to define such a theory in what follows.

We assume that gravity is quantized at some energy scale, $M_{QG}$
(the Planck scale,  or string scale),   that is at least an order of magnitude 
higher than  the UV scale, $M_{SM}$,  of  non-gravitational  interactions, loosely  referred below 
as Standard Model (SM) interactions.   $M_{SM}$ could be a scale 
at which the SM interactions themselves become UV complete, for instance, 
by grand-unifying into an asymptotically free theory.  In such a setup it's not unnatural 
to have two orders of magnitude hierarchy between  $M_{QG}$  and  
$M_{SM}$;   if so,  then gravity should be  well approximated by a 
classical field theory  below  the energy scale $M_{SM}$.  
At  these "low energies"  the  path integral can be  defined with all  the SM fields  
quantized using ${\hbar}$, while treating  gravity as an external classical 
field, pending specification of the rules of quantization for gravity.  The latter 
should give rise to further tiny corrections to already quantized SM 
processes  (see below).

Thus, at low energies the path integral  for quantized SM interactions reads as follows:
\beq
Z(g, J_n)  = {\rm const}\, \int d\mu ({\tilde \psi}_n) {\rm exp}\left (i \int d^4 x\sqrt{g} \left 
(  {\cal L} (g, {\tilde \psi}_n)   + J_n {\tilde \psi}_n  \right ) \right )\,,
\label{Z}
\eeq 
where $d\mu({\tilde \psi}_n)$ is a measure for  all the SM 
fields ${ \tilde \psi_n}$, that appropriately modes out gauge equivalent classes.  
The metric  field  $g$ is an external field, and so are the sources,  $J_n$'s, 
introduced for every single SM field. Then, the effective Lagrangian 
$L(g, \psi_n)$  used in (\ref {Sbar}) is defined 
as a Legendre transform of $ W(g, J_n) =- i {\rm ln} Z(g, J_n)$
\beq
\int d^4 x \sqrt {g} L  (g,  \psi_n)  \equiv   W (g, J_n) - \int d^4 x \sqrt {g} J_n \psi_n,  
\label{legandre}
\eeq
where  $\sqrt{g} \psi_n  \equiv  -i  \delta {\rm ln} Z(g, J_n)/\delta J_n$, is  $\sqrt g$ times 
the  vacuum expectation value of the SM field $\tilde \psi_n$,  in the presence of a source $J_n$.  
The obtained quantum  effective action  (\ref {legandre})  is a  1PI action.
Thus, all the quantum corrections due to non-gravitational interactions are
already taken into account in the  effective Lagrangian $L$. This Lagrangian is then inserted  
into  (\ref {Sbar}) to account for  dynamical gravity.    Note that the 
effective quantum Lagrangian $L(g,\psi_n)$ depends on the  classical fields, $g$ and $\psi_n$'s,  
only. The difference between these two sets  of fields is that the quantum corrections due 
to the SM interaction are already accounted for in the action for $\psi_n$'s, 
while the gravity quantum loops have not been taken into consideration yet.  
In what follows we will find it helpful to define an effective 
generating functional 
\beq
Z_{\rm SM} ( g, \psi_n) \equiv {\rm exp} \left ( i  \int d^4 x \sqrt {g} L  (g,  \psi_n)     \right )\, =
Z(g,J_n) {\rm exp} \left (   - i \int d^4 x \sqrt{g} J_n\psi_n  \right ),
\label{ZSM}
\eeq
that  includes all the SM loops, but does not include  quantized gravity.

In the end,  $g_{\mu\nu}$  in  (\ref {Sbar}) should also be quantized.
The corresponding quantum effects  are likely to become  of order one  
at scales $M_{QG}$, and they should be taken care of by a putative 
UV completion of the theory,  presumably via  new degrees of freedom that could 
appear at energies $\sim  M_{QG}$. These considerations can  
be postponed  for  a UV complete theory of gravity,  
such as string theory,  perhaps along the  lines proposed in \cite {Tseytlin}.   
However, there is an immediate  issue, 
irrespective of the form of UV completion. It concerns the  low-energy  effective theory:  
the quantum gravity corrections should  be  small  at scales well below  $M_{GQ}$ 
for our approximations above to  be  meaningful. For instance, in Einstein's gravity,  supplied 
with a diff-invariant UV cutoff  for gravity loops (that requires additional counter-terms to 
retain diff invariance), one generates  higher dimensional operators that make  
negligible contributions at energies below $M_{QG}$.  In the present case, however,  
one first needs to define  the rules   of calculation of the gravity loops given that the 
classical  action (\ref {Sbar})  has an unusual form. To define these rules, and 
check whether  gravity loop corrections are small, is  our goal in the reminder 
of this section.  

We  define an {\it extended} action: 
\beq
{\bar S}_{q,\lambda} = {1\over q}   \int d^4 x \sqrt{g} \left ( {1\over2 } R + L \right ) 
+ \lambda (V_g -q) \,,
\label{qlambda}
\eeq
and  write  down the  path integral for gravity as follows
\beq
Z_g   = {\rm const} \int d\mu(g)\, dq \,d \lambda \, {\rm exp} ( i  {\bar S}_{q,\lambda} )\,,
\label{zg}
\eeq
where $d \mu (g) $ is a  measure over diff-inequivalent metric fields. Note, that 
the fields of the 1PI SM action, $ \psi_n$'s,  play a  role  of  
external fields in the path integral for gravity.  Furthermore, 
one also integrates w.r.t. the {\it parameters} $q$ and $\lambda$ in this path integral.

The  expression in (\ref {zg})  can be rewritten in terms of the  path integral for the SM fields
$Z_{\rm SM}$  given in  (\ref {ZSM}):
\beq
Z_g  = {\rm const} \int d\mu(g)\, dq \,d \lambda \, \left (   {e}^{ iS_{\rm EH} }
\, Z_{\rm SM} ( g, \psi_n)  \right)^{1\over q}\, e^{i \lambda (V_g -q) }\, ,
\label{zgtot}
\eeq
where $S_{\rm EH}$  is the Einstein-Hilbert action for gravity. 
The above path integral defines an algorithm for calculating quantum corrections -- 
both  due to the SM interactions and gravity:  The SM loops are done in a conventional way,  
assuming the metric  to be an external classical field;  this gives rise  to  $Z_{\rm SM}  ( g, \psi_n)$.  
Furthermore,  for calculation  of gravity loops one  is invited to use an unconventional 
prescription  specified  either by (\ref {zgtot}),  or equivalently,  by (\ref {zg}).  

In this proposal, the parameter $q$ may be regarded as a second   
Planck's constant that governs the gravity loops (recall that  SM loops governed by the standard $\hbar$ 
are already taken into account in (\ref {zg})).  Furthermore, 
one integrates w.r.t. the second Planck's constant, however, the value of the 
latter is also constrained by the value of the invariant  
four-volume due to integration w.r.t. $\lambda$. The form of the extended action  
(\ref {qlambda}), unlike  that in (\ref {Sbar}),   is useful for thinking  of  the formulation of 
the path integral,  or canonical momenta  and Hamiltonian of the theory\footnote{The idea of integration 
w.r.t. the parameters is adopted from \cite {Tseytlin},  although  the path integral  
here, and its interpretation, differ  somewhat from that in \cite {Tseytlin}.}.
Having the path integral set up in (\ref {zg}) ,  we can  integrate  w.r.t. $q$ and $\lambda$ 
that would give rise to   
\beq
Z_g  = {\rm const} \int d\mu(g)\,  \, {\rm exp} ( i {\bar S} )\,,
\label{zg1}
\eeq
with $\bar S$ defined in (\ref {Sbar}). 
     
The trouble with the gravity loops in the effective field theory approach,  
can be understood  either in the language of (\ref {zgtot})  or  of (\ref {zg1}).
The latter presentation is shorter, so we reiterate it here from \cite {Tseytlin}
by observing that the $1/V_g$  factor in (\ref {Sbar}) is  rescaling  what would have been 
the Planck's constant for the gravity loops in a conventional effective field theory 
approach to Einstein's gravity;  that is,   we should take all the quantum gravity loop 
corrections  calculated in the conventional approach and make a replacement,  
$\hbar \to \hbar  V_g$ \cite {Tseytlin}.  Adopting this   procedure for the gravity loops, one  would get: 
\beq
{\bar S}_{Ren} = {1 \over V_g} \int d^4 x  \sqrt {g} \left ( {1 \over 2} R  
+  L  (g, \psi_n)  +   V_g L_1 (g, \psi_n)+ V_g^2 L_2 (g, \psi_n) +... \right )\,,
\label{SbarRen}
\eeq 
where $L_1,L_2, ..$ contain all possible terms consistent with diffeomorphism
and SM internal symmetries. The gravity loop  corrections are huge, 
since $V_g$ is huge. The new terms  ruin  the  above-presented solution 
of the cosmological constant problem.

It should be noted,  that there is yet another class of loop corrections  
if one quantizes  graviton fluctuations in the theory (\ref {Sbar}) on a given background 
solution\footnote{I thank Arkady Tseytlin for bringing  this point to my attention.}.
To consider the effects of these fluctuations, let us decompose the metric   
as a background  and fluctuation, schematically, $g =g_b +h$, where $h$ is being 
treated as small. Then, the inverse volume factor, $V_g^{-1}$, multiplying the  action $S$ in (\ref {Sbar}), 
can also be expanded  as follows: $V_g^{-1} = V_b^{-1} - V_b^{-2} H_h+...$, where $V_b = \int d^4 x\sqrt{g_b}$  and 
$H_h \equiv  \int d^4 x\sqrt{g_b}{\rm tr}(g_b^{-1} h)/2$, and so on.  It is clear that the  term,   $- V_b^{-2} H_h$
(and all the other subsequent terms containing higher powers of $h$),
will produce new unconventional interaction vertices when Wick-contracted with powers of  $h$ in 
the expansion of the action $S$.   While an extra effort would be required to work out  all 
these unconventional vertices, one should point out that all of them will be suppressed by 
powers of the background volume, $V_b$. Indeed, in the expression $- V_b^{-2} H_h$ one power
of the inverse volume $V^{-1}_b$ will be spent to offset the volume factor in the expression  $H_h$, while  the second 
power of $V^{-1}_b$ will be suppressing the fluctuations of $h$.  Thus, all the new vertices 
will come suppressed by  as many powers of $V^{-1}_b$ as the power of the fluctuation $h$ 
arising from the expansion of $V_g^{-1}$ in (\ref {Sbar}).  This suggests that the loop 
effects discussed in the present  paragraph could be  assumed to be small and be neglected.   
Similar considerations  apply to the proposal  discussed in  the next section.

\vskip 0.2cm

{\it 4.  Dealing with the problems:}  To avoid the  difficulty with the quantum loops discussed 
in the previous section,  let us  introduce the following  action  instead of (\ref {Sbar}): 
\beq
{A} = {  V_f \over V_g} S  +   \int d^D y \sqrt{f}\,  \left (  {M_f^{D-2}\over 2 }  \,R (y) + c_0 M^D  \cdots  \right )   \,,
\label{A}
\eeq 
where  a  second metric $f_{AB}(y), ~A,B =0,1,2,3,..D-1$, 
has been used,  and $V_f = \int d^Dy \sqrt{ f(y)}$,  in the $\mpl=1$ units.
Note that while the action $S$ defined in (\ref {action})  is four dimensional, the $f$-metric 
could  live in $D\ge 4$ dimensions in general.

The action of  the $f$-universe  has a certain vacuum energy scale 
$M $,  and a scale that determines the strength of its  gravitational coupling is $M_f$.
Depending on details of the theory  -- encoded in the dots in (\ref {A}) --  there may or may not be 
a stable hierarchy between the scales $M_f$ and $M$ (see below). 

The main idea is that  in (\ref {A}) any shift of $L$  by a constant,  $L\to L+c$, 
converts $c$ into a cosmological constant of the  $f$-universe,  thus removing it from 
the $g$-universe,  where we presumably reside.   Therefore, while the curvature in our 
universe is (nearly) zero, the other universe could  be highly curved.  

An analog of the {\it extended} action (\ref {qlambda}) 
now takes the form:
\beq
{A}_{q,\lambda} = {1\over q}   \int d^4 x \sqrt{g} \left ( {1\over2 } R + L \right ) 
+ \lambda \left ( {V_g\over V_f} -q \right )  +  
\int d^D y \sqrt{f}\,  \left (  {M_f^{D-2}\over 2 }  \,R (y) + c_0 M^D  \right ).
\label{qlambda_fg}
\eeq
This can be used  to define the path integral that includes integration w.r.t. 
$q$ and $\lambda$,  as discussed  in detail in the previous section.
Since all the essential steps of that construction carry through with  a straightforward 
extension to include the dynamics of the second metric $f$,  we will not repeat them here\footnote
{As it's evident from the above, $f$ is quantized  in a conventional way with $\hbar$.}.
Furthermore, in what follows we will use, for brevity,  the form of the action (\ref {A}),
obtained from the extended action (\ref {qlambda_fg})  by integrating out $q$ and $\lambda$.

In order for the gravity loops not to ruin the crucial classical property of the action, 
one should make sure that $V_f>>V_g$: then,  the rescaling of what would have been 
the Planck's constant  for gravity loops in a conventional approach is $\hbar \to \hbar  (V_g/V_f)$,   and the action including the gravity loop corrections 
would take the  form 
\beq
{  V_f \over V_g} \left [  {1\over 2} R  
+  L  +   {V_g\over V_f} L_1 
\cdots   \right ]  +   
\int d^D y \sqrt{f}\, \left (   {M_f^{D-2}\over 2 }  \,R   + c_0 M^D + c_1 R^2  +c_2 {{\bar S}^2 \over M_f^D} \cdots  \right ).
\label{Aren}
\eeq 
As long as  $V_f>>V_g$,  all the corrections proportional to $V_g/V_f$ can  be neglected. 
There are  also terms similar to the ones discussed in the last paragraph on
the previous section, but they are harmless for the same reasons as  before\footnote{Thus, the  
gravity loop corrections  to both gravity itself,  and the standard model processes, 
either vanish  or are very small in this prescription. However,   the theory still needs 
UV completion to make sense of its unusual form for $g$ and $f$ gravities.  
I thank David Pirtskhalava for very useful discussions on these points.}.  
This is not all however,  the gravity loop diagrams in the $f$-universe  generate  two 
groups  of new terms -- first, the terms containing higher powers and derivatives  of  curvatures $R(f)'s$,
and second,  terms containing  powers of  ${\bar S}$ (and their products 
with  powers of the $R$'s and derivatives); some of these terms are 
displayed in  (\ref {Aren}).   All these terms, however,  introduce small corrections, as  
it will be clear  from the discussion  on the hierarchy between 
the scales  in the $g$- and $f$-universes given below.

In general,  both $V_f$ and $V_g$  are  divergent. It is sufficient  for our purposes that 
the condition $V_f/V_g >>1$ is satisfied, even 
though $V_f$ and $V_g$ individually  tend to infinity\footnote{As  emphasized  
above, we  assume that these infinities are first regularized, and the regularization 
is removed at the end of calculations.}.  
For  considerations  of the ratio,  $V_f/V_g$,  it is  convenient 
to invoke the Euclidean  space  to get a sense of  the ratio  
of the Euclidean four-volumes, $V_f/V_g$, as will be  done 
below.

Then, how do we achieve  the condition  $V_f/V_g>>1$? 
To fulfill this  we're going to  explore  technically natural hierarchies  between 
parameters of the theory. First of all, we assume 
that the $g$-universe has supersymmetry broken at some high scale, and therefore,  there is a 
natural value of its vacuum energy density proportional to  $E^4_{vac}$.  The  
scale $E_{vac}$  can  be anywhere  between a few $TeV$  and the GUT scale,  $\mu_{GUT} \sim 10^{16}\,GeV$.
As to the $f$-universe,  it's presumably uncontroversial  to set  $M_f\sim \mpl$,  but  also we'd need 
the scale  $M$  to be somewhat higher than $E_{vac}$.   The latter condition should  be 
natural, since  without special arrangements one would expect  $M \sim M_f\sim \mpl$,
and since $E_{vac} << \mpl$, one would  also get  $E_{vac} < M$.  If so, then 
the vacuum energy of the $g$-universe, $E^4_{vac}$, would  make  a small 
contribution to the pre-existing vacuum energy of the $f$-universe. 
In short,  the vacuum energy density of the $f$-universe, $c_0M^D$, 
would dominate over the vacuum energy density that gets delegated 
to the $f$-universe, from the $g$-universe .

While one could try to explore a case when the $f$-universe has a 
positive vacuum energy density,  it seems more straightforward  to make  a mild assumption  
that the curvature  due to the term $c_0 M^D$ in the $f$-universe  is  negative (AdS like) 
to begin with. In that case,  the $f$-universe can be exactly supersymmetric, described by 
an unbroken  supergravity.

For instance, if we were to consider $D=4$, the $f$-universe could be described 
by supergravity  with the "Planck scale" equal to  $M_f$,  and  the quantity 
\beq
3{\bar \lambda}^2  \equiv  { {\bar S}+c_0M^4 } \,,
\label{l}
\eeq
acting as  its vacuum energy  density.  The action (\ref {A})  completed to  the one of the 
$N=1$ AdS supergravity \cite {Townsend}  
would then  be written  as: 
\beq
A_{SUGRA} = \int d^4 y \, {\tilde e}  \left (  {M_f^2\over 2} R({\tilde e}, {\tilde \omega}) - 
\epsilon^{\mu\nu\alpha\beta} {\bar \psi}_\mu \gamma_5 \gamma_\nu D_\alpha \psi_\beta
+ 3 {\bar  \lambda}^2    -  {2 {\bar \lambda} \over M_f} {\bar \psi}_\mu \sigma^{\mu\nu}  \psi_\nu
\right )\,,
\label{sugra}
\eeq
where $\tilde e$ is the determinant of the vierbein of the $f$-metric,  $\tilde \omega$ is its spin connection,
$D = \partial -{1\over 2} {\tilde \omega} \sigma $ is the covariant derivative,  and $\psi_\mu$ is the 
Rarita-Schwinger  field  describing a $f$-gravitino.  While  a supergravity action is not the only one that 
can  help reach our goal,  the motivation  to  consider it can perhaps be attributed to the fact that 
supergravities naturally emerge in the  low-energy limit  of superstrings.  

The quantity $\bar S$  enters into ${\bar \lambda}$ in (\ref {sugra}),  while the latter defines
the  cosmological constant  (with AdS sign) as well as a quadratic term for the gravitino
\footnote{This is certainly not a gravitino mass term  \cite {DeserZumino}.}.
Thus, the entire $g$-universe enters this action via the parameter $\bar \lambda$  defined in 
(\ref {l}).  The gravitino  bilinear term in (\ref {sugra})  would also give a nonzero contribution into 
the equation of motion for the metric  $g$, however,  the respective new term will  be proportional to 
the  gravitino bilinear, which is zero on classical solutions. Thus this term will not change 
our conclusions on the cosmological constant\footnote{For  ${\bar \lambda}^2$ to be positive the scale $M$ should be  
(somewhat) higher than the scale $E_{vac}$. This could  be arranged  without any fine tunings
as discussed above.}. 

There is no reason for the parameter $\bar \lambda$ to be much smaller than $M^2_f$;  quantum corrections 
would  renormalize the former up to the scale of the latter,   even if we started  with a  large 
hierarchy between them.  On the other hand, we do need
some small hierarchy between $M_f$ and ${\bar \lambda}^{1/2}$, essentially to be sure that 
AdS curvature of the $f$-universe can reliably be described  in the supergravity approximation.
For this, an order of magnitude hierarchy, $M_f \sim 10 {\bar \lambda}^{1/2}$,   would be 
more than sufficient. While this hierarchy  could perhaps be attributed,  without too much  of anxiety,
to  the $4 \pi^2$ loop factor's arising at various places,  we note that it could  be generated dynamically 
if we were to introduce  more general supersymmetric theory with some matter fields in 
the   Lagrangian:  the  $N_m$ matter 
fields  with a characteristic scale $M_m$ would renormalize additively the Planck  scale $M_f$   
via  the Adler-Zee mechanism producing, $M_f^2 \to M_f^2 + \Delta M_f^2$, where  
$\Delta M_f^2 \sim  N_m M^2_m$  \cite {Adler}, \cite {Zee}, while  renormalisation of the cosmological 
constant $\bar \lambda$ due to the complete SUSY multiplets of matter  would have  been zero.   Thus,  
we could adopt,  $M_f \sim 10 {\bar \lambda}^{1/2}$, as a technically natural choice.
If so, then the hierarchies  $\mpl \sim M_f  \sim 10  M$,  $M \gsim 10  E_{vac}$, ensure
that all the corrections in (\ref {Aren}) are negligible in comparison  
with the terms in (\ref {A}).

Having the scales clarified, let us see how this plays out  for the cosmological constant for 
a general $D$-dimensional $f$-universe.  
First  we consider the case when  $f$ is not   among the  fields  $\psi_n, n=0,1,2,3,.. $. 
Then, the new terms  in (\ref {A}) or (\ref {sugra})  do not affect  the equations (\ref {Tseytlin}), except that 
they introduce  a  overall multiplier $V_f$.  Thus,  the  cosmological  constant is eliminated from 
the $g$-universe.  There is, however, a new equation due to variation w.r.t. $f$:
\beq
M_f^{D-2} (R_{AB} (y)  - {1\over 2} f_{AB} R(y)) =  f_{AB}  ( {\bar S} + c_0 M^D) +\cdots  \,.  
\label{Eqf}
\eeq
The right hand side contains the vacuum energy  generated in our universe, 
${\bar S}  =  {[ E_{vac}^4]   \over V_g} =  E_{vac}^4$,  
as well as that  of the  $f$-universe.  According to our construction,   the net energy density
is negative, so that  the $f$-universe  has an AdS curvature. If so, then $V_f = \infty$ even in Euclidean space.  
Then,  to reach our goal it  is  sufficient to have Euclidean $V_g$ finite, so that  $V_f>>V_g$.  
A de Sitter  universe with  Euclidean $V_g =H_0^{-4}$ would fit the data and satisfy the 
above criterium\footnote{That $V_f^{AdS}/V_g^{dS} \to \infty$ can also be seen in Lorentzian signature, 
by calculating the ratio, e.g., in the global coordinate systems,  for the universal covering 
of AdS,  and the dS space.}. However,  the entire cosmological   constant has  been eliminated 
from the $g$-universe, and thus it's not easy any more to  get $V_g =H_0^{-4}$.  We'll  
discuss below how this could nevertheless be achieved.

\vskip 0.2cm

{\it 5. Getting the accelerated universe:}  One needs to get a dS metric in the 
$g$-universe without using a vacuum  energy or a scalar potential.  More precisely, 
one would need to get the  small  dS curvature due to the terms in 
the Lagrangian  (\ref {L}) that explicitly depend on  $g$. 

There might be a few  ways of achieving that: e.g., by  invoking Lorentz invariant condensates 
of some vector fields with a coherence length comparable with $H_0^{-1}$, or by 
using field theories with higher derivatives but no Ostrogradsky instabilities.  
Such proposals could produce  dark energy due to  terms  that aren't  potentials, but depend   
on the metric $g$, so that the last term on the r.h.s. of the trace equation (\ref {Tseytlin}), 
would define the cosmic speed-up.  
 
We briefly comment here on a possibility to obtain  this feature  due 
to massive gravity.   Nonlinear massive gravity \cite {dRG,dRGT}, or some of its extensions
\cite {Rosen,Quasidilaton,ExtendedQD,RampeiDavid},  introduce  graviton mass $m$ as a small 
parameter, $m\sim H_0$, in a technically  natural way  \cite {DavidNonrenorm}; these  theories  also 
produce  self-accelerated solutions with a dS background 
\cite {SA};   moreover, the fluctuations on these backgrounds are healthy when the pure massive 
graviton is amended with a dilaton-like field \cite {ExtendedQD,RampeiDavid} 
(for theory reviews  of  massive gravity see,  \cite {Kurt,Claudia}).

Let us briefly outline  how massive gravity would produce $R\sim m^2$ in 
the trace equation (\ref {Tseytlin}). For this we  put aside  the matter Lagrangian and assume that 
$L$  represents instead the  diffeomorphism invariant potential of massive gravity \cite {dRGT}:
\beq
L = \mpl^2 m^2 U(K) = \mpl^2 m^2(det_2(K)+ \alpha_3 det_3(K) + \alpha_4 det_4(K)), ~~~ 
K = 1 - \sqrt {g^{-1} \gamma}\,,
\label{mGR}   
\eeq
where,  the matrix $K$   is defined via an inverse  of the metric 
$g$  and  a fiducial  metric $\gamma$;   we chose $\gamma$ to be  a metric of Minkowski 
space,  $\gamma_{\mu\nu} = \partial_\mu \phi^a \partial_\nu \phi^b \eta_{ab}$,  written in an arbitrary 
coordinate system parametrized by $\phi^a, a=0,1,2,3$. The  $\phi^a (x)$  fields
also represent the St\"uckelberg fields that guarantee  diffeomorphism  invariance of 
(\ref {mGR}). The square root of a matrix  and its traces are defined via its eigenvalues, and 
$\alpha_3,\alpha_4$, are some free parameters.   Note that all possible values of the 
three parameters of the theory, $m, \alpha_2, \alpha_3$, are technically natural \cite {DavidNonrenorm}. 
Furthermore, the quasidilaton is introduced by  requiring that the rescaling of the $\phi^a$ coordinates
w.r.t. the $x^\mu$ coordinates  be promoted into a global symmetry; this amounts to adding into 
(\ref {mGR}) the kinetic term for the quasidilaton $\sigma$ (and possibly  some other derivative terms \cite 
{ExtendedQD}),   and replacing $\gamma \to e^{2\sigma/\mpl} \gamma $.

Let us now look at the trace equation  in (\ref {Tseytlin}): the trace of the stress-tensor, call it  $T^{g}$,  
is  obtained by the standard variation of  $[\sqrt {g} L] = \mpl^2 m^2 [\sqrt {g} U]$.
On the self-accelerated solutions  this  trace equals to a constant,  $T^{g} \sim \mpl^2 m^2$. Therefore,
$T^{g}$ in the l.h.s. of  (\ref {Tseytlin}) will cancel  with  $\langle T^{g} \rangle $ on the r.h.s.; the remaining  
trace  equation will take the form
\beq
R =   - 2 m^2  \langle g^{\mu\nu}   {\partial U(K) \over \partial g^{\mu\nu}}  \rangle\,.
\label{RmGR}
\eeq
On the selfaccelerated solutions, however,  $g^{\mu\nu}   {\partial U(K) / \partial g^{\mu\nu}}|_{SA} =
-C(\alpha_2,\alpha_3)$,  is also a constant, that  depends on the parameters $\alpha_2$  and $\alpha_3$. 
Therefore, its  average yields the same constant, and we get $R = 2 m^2  C(\alpha_2,\alpha_3)$. For 
a certain reasonable magnitudes,  and certain signs of the parameters, one gets  
the dS curvature  of the order,  $m^2 \sim H_0^2$, in a technically 
natural way.   Quasidilaton does not change this conclusion, it only affects (improves)   
dynamics of small perturbations above the solution \cite {RampeiDavid}.  
Thus, to summarize, the above approach enables to remove the big cosmological constant,  
and to get a small  space-time curvature determined by the graviton mass.

In the approach adopted above $\gamma$  was  taken to be independent of 
the $f$-metric,  that was used to remove the big cosmological constant. 
We've discussed  the case  when  $f$ was an  AdS metric,  while $\gamma $ was flat. 
However, neither  of these choices  are ordained -- we only require that space-time 
described by $f$ to have an infinite  Euclidean volume. It is intriguing, therefore,   to consider $\gamma$  
to be related (perhaps identified?)  with $f$.  In that case, $\gamma$ cannot be fixed a priori, but will be 
determined by the  $f$  equation of motion (\ref {Eqf}); the latter will now   be modified 
due to the terms in (\ref {mGR}), but  the modification is proportional to $m^2\sim H_0^2 << M^2_f$, 
and should be negligible.  If such a  framework can be made to work in detail, 
this would provide an additional arguments for amending  Tseytlin's approach  by 
the $f$-metric, and conversely,  would introduce an out-of-our-universe 
dynamics for the fiducial metric of massive gravity. 

On a more sobering note, massive gravity and its extensions are strongly coupled theories 
at energies way below $\mpl$; while  this may not be in conflict with observations in  our 
universe  due to the Vainshtein mechanism \cite {Vainshtein}  in its intricate cosmological  and 
astrophysical form \cite {VainshteinDeffayet}, \cite {MassiveCosmology}, \cite{GigaLasha}, nevertheless,  
it still remains to be understood how to go above the strong scale,  and show that superluminal 
phase and group velocities obtained on certain backgrounds probing this strong scale, 
are indeed artifacts to be removed in a complete treatment.

\vskip 0.2cm

{\it 6. Conclusions and outlook:} The proposed approach  eliminates  the  cosmological constant,
at least in a simple setup where  there are a few (non-proliferating) vacua  with well-separated 
hierarchy between their energy densities,  and allowed transitions between them. What is eliminated  is 
what would have been an asymptotic future value of the cosmological constant for such a potential 
in GR; for instance, for two vacua, "false" and "true",  with allowed transitions from "false" to "true",  
the "true" vacuum energy is eliminated. This is  similar to the proposal of \cite {ADDG}, but here 
the action functional  is available and it is stable w.r.t. quantum loop corrections, including loops 
of gravity in an effective  field theory approach. 

The dark energy component can be introduced  via the  Lorentz invariant 
condensates of vector fields, or via derivatively interacting scalar fields.  
We  briefly discussed   how the accelerated universe  
could be due to massive gravity in this approach.

The proposed scheme is rather unusual, as it involves nonlocal 
terms in otherwise local Einstein's equations, making  it difficult  to  be  
satisfied with this aspect. However, the cosmological constant problem is a  
long-standing  enigma of a tremendous magnitude, 
and any insight into its possible dynamical solution within the well-defined rules of the  low-energy 
field theory approach, is extremely  important, and should be welcomed.  

As an outlook, just three  comments  on the literature:  

Ref. \cite {ADDG} has made arguments for a connection of  the  "high-pass filter"  modification of gravity
with a specific theory containing the averages $\langle \cdots \rangle$. It might be interesting
to see if the present proposal could also be connected to some "high-pass filter" 
modified gravities discussed in \cite {ADDG}. Conversely, one could then hope to find an action principle 
for the equations of \cite {ADDG}, and address the issue of the gravity loops for them.

The original motivation of Tseytlin was to obtain 
the unconventional action (\ref {Sbar})  by  including the winding modes 
of string theory. It would be interesting to see  if any proposal along this
idea can give the action (\ref {A}), or a version of it.

Refs. \cite {Kaloper} have recently  discussed  gravity 
equations  involving the averages $\langle \cdots \rangle$,  with the goal to sequester 
the Standard Model vacuum energy.  The equations  and physical picture  obtained  in \cite {Kaloper}  are 
different from the  ones discussed in the present work.  It is argued that 
the particle physics loops are under control in \cite {Kaloper}, while the 
gravity loops were not considered.  It  could  perhaps be interesting to apply
the proposal of the present work to  deal with the gravity 
loops in \cite {Kaloper}.

\vskip  0.5cm

{\it Acknowledgements:} I'd like to thank  Matt Kleban,  
Massimo Porrati, and especially David Pirtskhalava and  Arkady Tseytlin for  
valuable  communications. The work is supported by NASA grant NNX12AF86G S06, 
and NSF grant  PHY-1316452.
 
 \vskip 0.2cm
 

\end{document}